# The Nature of the On-Street Parking Search

Aleksey Ogulenko, Itzhak Benenson*, Nir Fulman

Porter School of the Environment and Earth Sciences, Tel Aviv University

ogulenko.a.p@gmail.com, bennya@tauex.tau.ac.il, fulmannir@gmail.com

*corresponding author

## Abstract

Parking occupancy in a delineated area is defined by three major parameters – the rate of car arrivals, the dwell time of already parked cars, and the willingness of drivers to continue their search for a vacant parking spot. We investigate a series of theoretical and numeric models, both deterministic and stochastic, that describe parking dynamics in an area as dependent on these parameters, over the entire spectrum of the demand-to-supply ratio, focusing on the case when the demand is close to or above the supply.

We demonstrate that a simple deterministic model provides a good analytical approximation for the major characteristics of a parking system – the average fraction of cars among the arrivals that will find parking in the area, the average number of cars that cruise for parking, and the average cruising time. Stochastic models make it possible to estimate the distributions of these characteristics as well as the parameters defined by the distribution PDF, like the fraction of the arriving cars that find parking in less than $t$ minutes. The results are robust to the distribution of drivers' dwell and renege times and can be directly applied to assess the real-world parking dynamics.

# The Nature of the On-Street Parking Search

## 1. Introduction

The general view of urban parking considers drivers who arrive to perform activities in an area $A$, and have to find a nearby vacant parking spot. Parking occupancy and the price of parking in the vicinity of the destination define the driver's two major parking decisions – whether to start cruising for on-street parking in $A$ or go directly to a parking lot. In the case of unsuccessful cruising, the driver must decide whether to quit cruising and consider another solution, such as parking at the lot or canceling the planned activity altogether. If the parking search is successful, a car's dwell time is defined by the driver's activities in $A$, and by the cost of parking there.

The outcome of a driver's parking search is influenced by parking occupancy in the vicinity of the driver's destination. In the case of low occupancy, the driver successfully parks close to the destination. In the case of high occupancy, the search time and the distance between the potential parking spot and the destination become very uncertain. These high-occupancy conditions are characteristic of cities' business districts during the day and of overnight parking in dense residential areas.

Drivers' collective parking behavior defines the emerging parking pattern in $A$. In the long run, drivers who frequently have either a long search and/or a long walk to the destination may decide to forsake cruising and go directly to a parking lot on arriving to $A$. Furthermore, drivers for whom the overall losses of time and money due to a long parking search, a long walk time, or high parking fees may decide to quit using their car to reach $A$ and change either the destination of their activity or their transportation mode and use public transport to arrive to A.

The rough threshold that of the scarce parking conditions was proposed by Shoup (Shoup 2011) and is commonly accepted as the "85%" rule: As long as one of seven parking spots on every road link is vacant, parking is practically immediate. The 85% rule is highly intuitive - the average car occupies approximately 5m of the curb length and, thus, 85% or "6 of 7" occupancy results in a 30m average distance between the vacant parking spot and the destination. With an increase in occupancy above 85%, the average search time and distance to the destination grow, and the dependence of these characteristics on occupancy is essentially non-linear (Fulman and Benenson 2021; Levy and Benenson 2015; Levy et al. 2013).

Parking occupancy in $A$ is defined by three major parameters – the rate of car arrivals to $A$, the dwell time of cars parked there, and the willingness of drivers to continue their parking search when they have yet to find a vacant spot. In this paper, we investigate a series of theoretical models, deterministic and stochastic, that describe parking dynamics in $A$ as dependent on these parameters over the entire spectrum of demand-to-supply ratio. We confirm the analytical results with numeric simulation. Naturally, we focus on the case when the demand is close to or above the supply.

The rest of the paper is organized as follows: In section 2 we present the state of the art of parking search modeling; in section 3 we propose and investigate deterministic models of parking dynamics, the basic model and the model that explicitly accounts for the possible dependence of drivers' behavior on the accumulated search time. Section 4 is devoted to studying stochastic parking models that follow the same principles that were employed in section 3, and section 5 indicates how the investigated models can be applied in the heterogeneous urban environment. We discuss our results and their applicability to real-world parking situations in section 6.

## 2. The models of parking search and parking pattern dynamics

Parking space is a common resource and every city decides whether and how to charge residents and visitors for using it (Lin et al. 2017; Jioudi et al. 2019). The economics of parking, especially the optimal parking prices (Inci 2015; Shoup 2021) is, the most popular topic of recent studies, and Western cities are slowly evolving towards adaptive and dynamic parking pricing (Saharan et al. 2020). The standard parking practice is zone-based: the city is divided into a few parking zones, usually the city center and several rings around the center. Prices are higher in the center and lower on the periphery, and on-street parking is free or cheap for the zone's residents and more expensive for visitors. Parking lot prices are, typically, more variable and depend on whether ownership of the lot is municipal or private, and the attractiveness of the neighborhood for visitors.

The ratio of on-street and off-street prices is critical for the urban parking policy, and on-street prices can be both lower or higher than off-street prices (Auchincloss et al. 2015). Strategically, the goal of parking pricing is to reduce on-street occupancy to the 85% level. However, the elasticity of parking prices is relatively low (Lehner and Peer 2019), and to decrease the rate of private cars arriving to areas with high demand, like urban business districts, the prices should be high. In addition, the effects of parking prices can deteriorate over time (Alemi et al. 2018; Lee et al. 2017; Assemi et al. 2020; Millard-Ball et al. 2020; Hampshire et al. 2016).

When parking occupancy exceeds 85%, cruising time becomes an indicator of the balance between the arrival rate, dwell time, and willingness to continue the parking search. The distribution of the cruising time is a key to understanding drivers' reactions to urban parking policy and pricing. The dependence of cruising time on parking system parameters is investigated in several papers, all applying a queueing theory. Millard-Ball et al. (2014) established a queueing model to evaluate the relationship between occupancy and the amount of cruising vehicles when studying the impacts of the San Francisco SFPark adaptive parking pricing program. Dowling et al. (2017, 2018, 2020) consider parking search as a process that can be described by a network of interacting queues, each representing cars that want to park at a certain link, with cruising cars migrating between queues with respect to the network topology. They apply their model to estimate the share of cruising drivers in the neighborhood of Belltown, Seattle, WA, USA, and find that on the most demanded streets cruising drivers can comprise up to 50% of the traffic. Xiao et al (2018) employ a queueing model to predict future occupancy from historical data, and employ SFPark experiment data to demonstrate the accuracy of the prediction. Levy and Benenson (2015) propose a simple method for estimating parking occupancy in the case of spatially heterogeneous demand and supply, while Fulman and Benenson (2021) improve this method to estimate driver cruising time depending on the demand and supply in the vicinity of the driver's destination.

In this paper, we propose analytical, deterministic, and stochastic models of parking and compare their outputs with numeric simulations. Our basic assumption is that the chance to find a vacant parking spot does not depend on the time already spent on a parking search. In the case of stochastic models, this assumption is employed in the queueing model of parking proposed by Larson and Sasanuma (2010), who formalize the parking search as a "Service In Random Order" (SIRO) queue. We extend their model to the case of an arbitrary distribution of the dwell time and drivers' willingness to continue the parking search, and focus on the effects of important parameters of the parking system, like the fraction of drivers who cruise less than a certain time. We obtain an effective approximation for the distribution of the basic parameters that characterize the on-street parking system, like the number of cars that cruise for parking and cruising time.

## 3. Deterministic view of the parking dynamics

Let us consider a bounded neighborhood $A$ with $c$ parking spots and assume that $A$ is small enough to guarantee that drivers with the destination in $A$ would cover all of it when searching for parking. We consider the state of parking in $A$ at discrete time moments $t$. The number of parked cars $P(t)$ in $A$ is defined by the rate $\lambda$ of cars' arrivals to $A$ per time step, the departure rate $\mu$, per time step, of departing cars that were previously parked in $A$, and the rate $\alpha$, per time step that drivers renege in the case of an unsuccessful parking search. We aim to estimate the characteristics of the parking system in $A$ in its equilibrium state – the number of cruising cars, the probability to park, and others, as dependent on $\lambda$, $\mu$, and $\alpha$. The variables and parameters that we use in this paper are listed in Appendix 1.

### 3.1. Constant dwell and renege rates (basic deterministic model)

Let us start with the simplest case of constant rates $\lambda$, $\mu$, and $\alpha$ that we call below the *basic deterministic model* and consider the extreme case of $\alpha = 0$, that is, cars that arrive to $A$ may cruise for parking infinitely. The dynamics of the $P(t)$ is represented in this case by the equation:

$$P(t + 1) = (1 - \mu)P(t) + \lambda \qquad (1)$$

Equation (1) has a unique and globally stable equilibrium

$$P^* = \lambda/\mu \qquad (2)$$

Indeed, denoting $\Delta P(t) = P(t) - P^*$ we obtain

$$\Delta P(t + 1) = (1 - \mu)\Delta P(t) \qquad (3)$$

That is, $\Delta P(t)$ converges to zero for any positive $0 < \mu < 1$.

However, $A$'s parking capacity is limited. That is, model (1) reflects the reality only when the equilibrium occupancy does not exceed $c$, or $P^* = \lambda/\mu \leq c$. If $\lambda/\mu > c$, then the cars that fail to park remain in the system infinitely, and the overall number of cars in the system steadily grows, at a constant rate $\lambda - c\mu$ per time step.

To make the model applicable for $\lambda/\mu > c$, we have to assume that the renege rate $\alpha$ is positive. We also assume that (a) arriving cars join the queue of the cruising cars and (b) when a spot is vacated, one of the cruising cars is *randomly selected from this queue, irrespective of the accumulated cruising time*. To fully formalize the parking search process, we consider the following order of departing, reneging, arriving, and parking during the time interval between the moments $t$ and $t + 1$:

- $t \to t + 1/3$: fraction $\mu$ of occupied spots are vacated and fraction $\alpha$ of cruising cars renege
- $t + 1/3 \to t + 2/3$: $\lambda$ cars join the queue
- $t + 2/3 \to t + 1$: cruising cars are randomly selected from the queue to occupy vacated spots. The number of selected cars is a minimum of the number of vacant spots and a number of cruising cars.

Given $\mu$, $\lambda$, and $\alpha$, the state of parking in $A$ can be described by two variables – the number of occupied spots $P(t)$, as above, and the number of cruising cars $Q(t)$:

$$\text{if } \lambda/\mu \leq c, P(t+1) = (1-\mu)P(t) + \lambda, \quad Q(t+1) = 0 \qquad (4)$$

$$\text{if } \lambda/\mu > c, P(t+1) = c, \qquad\qquad Q(t+1) = (1-\alpha)Q(t) + \lambda - c\mu$$

In the case of $\lambda/\mu \leq c$, the equilibrium solution $(P^*, Q^*)$ of (3) is given by (2): $P^* = \lambda/\mu, Q^* = 0$, while for $\lambda/\mu > c$, the equilibrium solution is $P^* = c$ and

$$Q^* = \frac{\lambda - c\mu}{\alpha} \qquad (5)$$

The solution $(c, Q^*)$ for $\lambda/\mu > c$ is also globally stable: Denoting $\Delta Q(t) = Q(t) - Q^*$, we obtain

$$\Delta Q(t+1) = (1-\alpha)\Delta Q(t) \qquad (6)$$

That is, $\Delta Q(t)$ converges to zero for any positive $0 < \alpha < 1$.

We call system of equations (4) *the basic deterministic model*.

In what follows we present the characteristics of the parking system as dependent on the ratio $\rho$ of the number of arrivals to the number of departures per time step. We call it the "*Arrivals-to-Departures*" (AD) ratio and calculate it as $\rho = \lambda/(c\mu)$. The case of $\lambda/\mu \leq c$ corresponds to $\rho \leq 1$, while the case of $\lambda/\mu > c$ to $\rho > 1$. Note that $\rho$ is dimensionless.

The cars that arrive to the system either park or renege. Let us estimate the equilibrium fraction $f^*$ of cars that succeed to park, among all cars that arrive to the system. If $\rho \leq 1$, then $f^* = 1$ that is, all arriving cars will park immediately after arrival, irrespective of the value of $\alpha$. For $\rho > 1$, some of the cars park while some renege, at every time step. In the equilibrium state, the number of cars that park during the time period $T$ is $c\mu T$, while the number of arriving cars is $\lambda T$, that is, the fraction of cars that succeed to park is

$$f^* = \frac{c\mu T}{\lambda T} = \frac{c\mu}{\lambda} = 1/\rho \qquad (7)$$

The equilibrium fraction of cars that renege is equal to $1 - f^* = \frac{\lambda T - c\mu T}{\lambda T} = \frac{\lambda - c\mu}{\lambda} = \frac{\rho - 1}{\rho}$, and does not depend on $\alpha$.

Evidently, the cruising time $w_\kappa$ ("κ" from Greek "κρουαζιέρες" – cruising) for $\rho > 1$ is non-zero. To estimate $w_\kappa$ for $\rho > 1$, let us note that in the equilibrium state, the probability $\varepsilon$ that a cruising car will park equals the number of parking places $c\mu$ vacated at $t + 1/3$, divided by the number of cars that search for parking. The latter is established at $t + 2/3$ and is equal to $\lambda + (1 - \alpha)Q^*$. As a result, for $\rho > 1$

$$\varepsilon = \frac{c\mu}{\lambda + (1-\alpha)Q^*} = \frac{c\mu\alpha}{\lambda - c\mu + c\mu\alpha} = \frac{\alpha}{\rho + \alpha - 1} \qquad (8)$$

Evidently, for $\rho \leq 1, \varepsilon = 1$.

For the constant $\lambda$, $\mu$, and $\alpha$, the fraction of cars that will continue cruising is constant and equals, for $\rho > 1$, to $\phi = (1 - \varepsilon)(1 - \alpha)$. Note that we split the transition from $t$ to $t + 1$ into three steps and $\phi$ describes the state of the system at $t + 1/3$, after all departures have happened and the cruising drivers have decided whether to continue cruising or renege. This defines the geometric distribution of the cars' cruising time with the average

$$w_\kappa = \sum_{i=0}^{\infty} i(1-\phi)\phi^i = \frac{\phi}{1-\phi} = \frac{(1-\varepsilon)(1-\alpha)}{\varepsilon + \alpha - \varepsilon\alpha} \qquad (9)$$

As should be expected, with the increase of $\rho$, $w_\kappa$ converges to $\frac{1}{\alpha} - 1$, one time step less than the average renege time, because we count the number of failures *until* the first success.

The driver may reconsider the decision to arrive to the area today if the cruising time yesterday was too long. In this respect, the fraction $\psi_\tau$ among all the cars arriving in the system that find parking in $\tau$ time steps or less is:

$$\psi_\tau = \varepsilon \sum_{i=0}^{\tau} \varphi^i = \varepsilon(1 - \phi^{\tau+1})/(1 - \phi) \qquad (10)$$

Figure 1 presents the equilibrium number $Q^*$ of cruising cars, the cruising cars' probability $\varepsilon$ to park, and the average cruising time $w_\kappa$ as dependent on $\rho$. In all examples below, the time step is equal to one minute and $\mu = \frac{1}{120}$, that is, the average parking time is set to 2 hours, while the values of $\alpha = \frac{1}{2}, \frac{1}{5}, \frac{1}{10}, \frac{1}{15}, \frac{1}{20}$ that is, the average renege time is set to 2, 5, 10, 15, and 20 minutes. For $\rho > 1$ the number of cruising cars $Q^* = \frac{c(\rho - 1)\mu}{\alpha}$ comprises a constant fraction $\frac{(\rho - 1)\mu}{\alpha}$ of the $A$'s capacity $c$ (Figure 1a), the probability to find parking decreases as $1/\rho$ (Figure 1b), and the average cruising time increases as $\frac{(1-\alpha)(\rho-1)}{\alpha\rho}$ (Figure 1c).

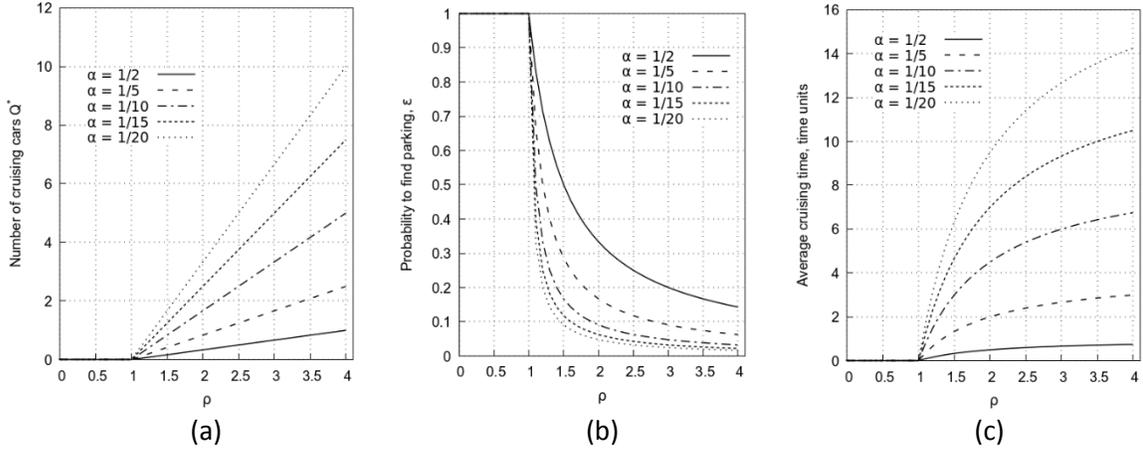

Fig. 1. Characteristics of the parking system in the equilibrium state of the basic deterministic model as dependent on the AD-ratio $\rho$, for $c = 20$, $\mu = \frac{1}{120}$, $\alpha = \frac{1}{2}, \frac{1}{5}, \frac{1}{10}, \frac{1}{15}, \frac{1}{20}$ and $\lambda = \rho c \mu$: (a) equilibrium number of cruising cars $Q^*$; (b) probability $\varepsilon$ for the cruising car to find parking; (c) average cruising time $w_\kappa$.

The constant departure and renege rates that we employed in this section result in a geometric distribution of the dwell and renege times - the probability to park for $s$ or more time steps is $(1 - \mu)^s$ and probability to cruise for $v$ or more time steps is $(1 - \alpha)^v$. These assumptions are qualitatively unrealistic: On the one hand, the probability $\mu$ to leave a spot immediately after parking and the non-zero probability $\alpha$ to quit cruising and leave the system immediately after arrival are, by definition, non-zero. On the other, dwell and cruising times can be arbitrarily large. To revoke these limitations, let us consider a deterministic model with the arbitrary distribution of the dwell and renege times.

### 3.2. Arbitrary distribution of dwell and renege time (age-structured model)

In this section, we establish a deterministic model for the arbitrary distributions of the dwell and cruising time, and assume that the arrival rate $\lambda$ is constant. The model of this section explicitly accounts for the accumulated cruising and dwell time of the car, or the cruising "age" of a car. By analogy to the population dynamics models, we call it the *age-structured* model of parking.

Let $\delta_s$ be the fraction of drivers who park for exactly $s$ time steps and $s_{max}$ be the maximal dwell time, $\sum_{i=1}^{s_{max}} \delta_i = 1$. The average dwell time is, then, $w_d = \sum_{i=1}^{s_{max}} i\delta_i$. Note that a constant departure probability $\mu$ results in the geometric distribution of the dwell time, $\delta_s = \mu(1 - \mu)^{s-1}$, $s_{max} = \infty$, and the average dwell time $w_d = 1/\mu$. Similarly, if $\gamma_v$ is the probability that the driver reneges after exactly $v$ time units of cruising and $v_{max}$ is the maximal renege time, then $\sum_{i=1}^{v_{max}} \gamma_i = 1$ and the average renege time is $w_r = \sum_{i=1}^{v_{max}} i\gamma_i$. The constant renege probability $\alpha$ results in the geometric distribution of the renege time $\gamma_v = \alpha(1-\alpha)^{v-1}$, $v_{max} = \infty$ and the average renege time $w_r = 1/\alpha$.

Let us consider the arbitrary distribution of the dwell time $s$ and renege time $v$, but limit the dwell time $s$ by $s_{max} < \infty$ and the renege time by the $v_{max} < \infty$. The state of the parking system at moment $t$ can be represented by two vectors: $\bar{P}(t) = \left(P^0(t), \dots, P^s(t), \dots, P^{s_{max}}(t)\right)^T$, where $s = \overline{0, s_{max}}$ and $P^s(t)$ is the number of drivers who parked at $A$ at a time moment $t-s$, and $\bar{Q}(t) = \left(Q^0(t), \dots, Q^v(t), \dots, Q^{v_{max}}(t)\right)^T$, where $v = \overline{0, v_{max}}$ and $Q^v(t)$ is the number of drivers who entered $A$ at time moment $t-v$.

As in the previous section, we start with the case of the zero renege rate. In this case, the dynamics of parking in $A$ are described by the following system of equations:

$$P^0(t+1) = \lambda$$
$$P^1(t+1) = (1-\delta_1)P^0(t)$$
$$P^2(t+1) = (1-\delta_2/(1-\delta_1))P^1(t)$$
$$\dots$$
$$P^s(t+1) = (1-\delta_s/(1-\sum_{i=1}^{s-1}\delta_i))P^{s-1}(t) \qquad (11)$$
$$\dots$$
$$P^{s_{max}}(t+1) = (1-\delta_{s_{max}}/(1-\sum_{i=1}^{s_{max}-1}\delta_i))P^{s_{max}-1}(t)$$

The solution of (11) converges to the steady equilibrium

$$\bar{P}^* = \left(\lambda, (1-\delta_1)\lambda, (1-\delta_1-\delta_2)\lambda, \dots, \left(1-\sum_{i=1}^{s_{max}-1}\delta_i\right)\lambda\right)^T \qquad (12)$$

and the value of $P(t) = \sum_{i=1}^{s_{max}} P^i(t)$ represents the total number of occupied spots in $A$.

Given $\sum_{i=1}^{s_{max}} \delta_i = 1$, one can easily demonstrate that the equilibrium value $P^*$ of $P(t)$ is:

$$P^* = \lambda w_d \qquad (13)$$

Formula (13) generalizes (2) for the case of arbitrary distribution of the dwell time $s$ and, as above, is valid only in the case of $P^* = \lambda w_d \leq c$.

In the case of $\lambda w_d > c$, all spots in $A$ are always occupied, the distribution of parking cars by their dwell time can be obtained from (12) substituting $\lambda = \lambda_c = c/w_d$, and some cars are always cruising for parking. To define the number of cruising cars and the age structure of the queue of cruising cars, we follow the assumptions about the sequence of departures, reneging, arrivals, and parking that are established in the previous section: Random choice of queueing cars for parking means that $c/w_d$ vacancies that are generated at the moment $t+1/3$ will be occupied by the cruising cars that are randomly selected at the moment $t+2/3$. The number of cruising cars at $t+2/3$ is $Q(t+2/3) = \sum_{i=1}^{v_{max}} Q^i(t) + \lambda$ and, thus, the probability of a cruising car parking between $t$ and $t+1$ is $\epsilon(t) = \frac{c}{w_d Q(t+2/3)}$. The equations that describe the dynamics of the queue of cruising cars are, thus, as follows:

$$Q^0(t+1) = \lambda(1-\epsilon(t))$$

$$Q^1(t+1) = (1-\gamma_1)(1-\epsilon(t))Q^0(t)$$

$$Q^2(t+1) = (1-\gamma_2/(1-\gamma_1))(1-\epsilon(t))Q^1(t)$$

...

$$Q^v(t+1) = (1-\gamma_v/(1-\sum_{i=1}^{v-1}\gamma_i))(1-\epsilon(t))Q^{v-1}(t) \qquad (14)$$

...

$$Q^{v_{max}}(t+1) = (1-\gamma_{v_{max}}/(1-\sum_{i=1}^{v_{max}-1}\gamma_i))(1-\epsilon(t))P^{v_{max}-1}(t)$$

Different from the case of exponential distribution of the renege time, the equilibrium solution $\overline{Q^*} = (Q^{1,*}, ..., Q^{v,*}, ..., Q^{v_{max},*})^T$ of (14) cannot be presented in a closed analytical form because of the non-linear dependence of $\epsilon(t)$ on $Q(t)$. However, the convergence of (14) to $\overline{Q^*}$ is fast, and for the realistic distributions of $\gamma_v$, $\overline{Q^*}$ always stabilizes in a few tens of iterations, irrespective of initial conditions. Note that the estimate of $\overline{Q^*}$ is accompanied by the estimates of the equilibrium fraction of parked cars $\epsilon^*$ and the equilibrium number of cruising cars $Q^*$.

Based on the $\overline{Q^*}$, $Q^*$ and $\epsilon^*$, the average cruising time $w_\kappa$ for the model (14), can be estimated as

$$w_\kappa = (\sum_{i=1}^{v_{max}} i(1-\gamma_i/(1-\sum_{j=1}^{i-1}\gamma_j))\epsilon^* Q^{i,*})/Q^* \qquad (15)$$

The fraction $\psi_\tau$ of cars that find parking in τ time steps or less is, evidently,

$$\psi_\tau = \sum_{i=0}^{\tau} Q^i / Q^* \qquad (16)$$

We will exploit the age-structured models (9) – (12) when comparing the outcomes of the deterministic and stochastic models of parking that we present in the next section.

### 3.3. Numeric model of parking

The deterministic model (11) – (14) cannot be fully analyzed analytically and its equilibrium solution $\overline{Q^*}$ should be estimated numerically. Similarly, the stochastic model of parking that we investigate in the next section can be investigated only analytically in the case of constant departure and renege rates. To overcome the limitations of the analytical approach, for each version of the model we have established a parallel numeric version.

The numeric version of the deterministic model just repeats the calculations represented by (11) – (14). The numeric model of parking that corresponds to the stochastic model is different and implements *agent* technology that considers every arriving car as a single agent. Agents arrive to $A$, cruise for parking, renege or park, and, if parked, leave the system when the parking time is over. An agent can be in one of two *states* – *cruising* or *parked*. Agents are considered at discrete time steps and their arrivals, cruising, transitions from the state of cruising to the state of parking, and departures are defined by the rules that repeat the rules of the stochastic model. For example, to represent Poisson arrivals at the rate $\lambda$ per time step, the Poisson distribution with the parameter $\lambda$ is constructed and, at every time step anew, the number of newly arriving agents is established according to this distribution. If an agent is cruising, then it can renege with the probability that is defined by the accumulated cruising time according to an *a priori* distribution of the cruising time (which, in case of a constant probability to renege, is geometric). The same approach is applied to a parked agent to determine whether to continue parking or depart. The number of agents that park at time step $t$ is defined as a minimum of the number of vacant spots at $t$ and the number of cruising agents at $t$. The agents that succeed to park are chosen from the queue by applying the *selection sampling* algorithm (Knuth 1997).

The sequence of agents' arrivals, departures, and transitions between states in the numeric model repeats the sequence established in section 3.1: At the stage $t \rightarrow t + 1/3$ each of the parked agents is checked and either continues to dwell or departs, and each cruising agent is checked and either continues to cruise or reneges. At the stage $t + 1/3 \rightarrow t + 2/3$ new agents arrive and join the queue; at the stage $t + 2/3 \rightarrow t + 1$ some of the cruising agents are randomly chosen to occupy the vacant spots, thus changing their state from cruising to parked.

In what follows, when an analytical study of the model is possible, we present the analytical solution together with the results of the numeric simulations. The close match between analytical and numeric outcomes in all of these cases makes it possible to rely on the numeric solutions in case when an analytical analysis is not possible.

## 4. Stochastic view on the parking dynamics

Parking processes are inherently stochastic and our goal in this section is to construct a stochastic model that reflects the same general view on parking that was employed in the deterministic models. Importantly, the stochastic model has to reflect that the chance to park for a car that just arrived is the same as that for a car that was cruising for a long time. To create such a model, we start with a simpler one that assumes that the cars park in the same order as they arrive. This is the $M/M/c$ queueing model that assumes that Poisson arrivals of cars at $A$ generates a single queue with exponential reneging (Ancker and Cafarian 1963; Montazer-Haghighi et al. 1986) that is served by $c$ servers (parking spots) with exponential departures, and the order of queue service is *First In First Out* (FIFO).

Note that in the deterministic models above we did not attempt to interpret the FIFO order of service, and always assumed that the reneging and parking cars are randomly chosen from the set of cruising cars. The random choice of service in the queueing model is called *Service In Random Order* (SIRO) queue discipline. The SIRO queueing models were introduced by (Palm 1937) on the eve of queue theory development, and studied extensively in the 1960s (Kingman 1962; Takács 1962) in regards to the case of infinitely patient i.e., not reneging, customers. Recently, Brenbjerg et al. (2016) compared equilibrium behavior of the impatient customers deciding when to arrive for service, for the SIRO and FIFO queueing disciplines.

In this paper, we focus on the SIRO queueing model that directly repeats an intuitive view of the parking search and parking, and is implemented in the above deterministic models. A possible interpretation of the FIFO queue discipline could be the parking search of cars that join the tail of the conga line that follows the same closed cruising path (Arnott and Williams 2017). In this case, the chance to park for the car that is first in the queue is higher than those following it.

Despite the difference in the FIFO and SIRO queueing disciplines, the well-known robustness of queueing systems with Poisson input (Medhi 2002, section 6.10.3), ensures similarity of the average characteristics of the queue process with both disciplines. Below, we confirm this similarity. Namely, we present an analytical investigation of the FIFO model of parking and demonstrate a very close match (1) between the numeric and analytical results obtained with the FIFO model, and (2) between the average characteristics of the numeric FIFO and SIRO models. We also demonstrate that, as mentioned in (Medhi 2002), the tails of the distributions of the parking process characteristics are different in the FIFO and SIRO models. We focus on the SIRO outputs as reflecting our views of the parking process in this respect. In all applications below, the numeric model serves for estimating the stochastic models' characteristics under the assumptions that do not allow for the analytical study.

### 4.1. The FIFO queue with the reneging model of parking and its equilibrium state

As above, we consider parking area $A$ with $c$ parking spots and interpret cruising in $A$ until finding the vacant spot as waiting in the queue. Formally, we assume that:

(1) Cars' arrivals to $A$ are described by a Poisson process with parameter $\lambda$.
(2) If there is a vacant spot in $A$, the arriving car parks immediately. Otherwise, when $A$ is full, the arriving car joins the waiting queue.
(3) Cars' dwell time has an exponential distribution with parameter $\mu$.
(4) Drivers are impatient: the time a queueing car will wait for parking has an exponential distribution with parameter $\alpha$.

In the following, we will call this set of assumptions the *stochastic model*, specifying the order of queue service (FIFO or SIRO), if necessary. As above, we will present the results as dependent on the AD ratio $\rho = \lambda/c\mu$. Montazer-Haghighi et al. (1986) have applied a contextually close FIFO model to estimate the average number of cars in the system and the average number of cars that leave the system without parking. Our analysis provides the distributions of these and other characteristics of the FIFO system in an analytical form (see Appendix A2 for more details).

Let $p_n(t)$ denote the probability that there are $n$ cars either parked or waiting for a vacant spot in $A$. The Kolmogorov equations for the model are as follows:

$$\begin{cases} p_0'(t) = -\lambda p_0(t) + \mu p_1(t), \\ p_k'(t) = -(\lambda + k\mu)p_k(t) + \lambda p_{k-1}(t) + (k+1)\mu p_{k+1}(t), & 1 \leq k \leq c-1, \\ p_c'(t) = -(\lambda + c\mu)p_c(t) + \lambda p_{c-1}(t) + (c\mu + \alpha)p_{c+1}(t), \\ p_k'(t) = -(\lambda + c\mu + (k-c)\alpha)p_k(t) + \lambda p_{k-1}(t) + (c\mu + (k-c+1)\alpha)p_{k+1}(t), k > c \end{cases} \quad (17)$$

Assuming the steady-state distribution exists, the equilibrium probabilities $p_k$ are the solutions of the following system of equations.

$$\begin{cases} 0 = -\lambda p_0 + \mu p_1, \\ 0 = -(\lambda + k\mu)p_k + \lambda p_{k-1} + (k+1)\mu p_{k+1}, & 1 \leq k \leq c-1, \\ 0 = -(\lambda + c\mu)p_c + \lambda p_{c-1} + (c\mu + \alpha)p_{c+1}, \\ 0 = -(\lambda + c\mu + (k-c)\alpha)p_k + \lambda p_{k-1} + (c\mu + (k-c+1)\alpha)p_{k+1}, & k > c. \end{cases} \quad (18)$$

Based on the equilibrium probabilities, we can calculate the key characteristics of the parking system: the average number of cars in the system, the average length of the queue, and the average waiting time in the queue. Solving the system of difference equations (18), we obtain

$$p_k = \frac{1}{k!}\left(\frac{\lambda}{\mu}\right)^k p_0, \quad k \leq c,$$

$$p_k = \frac{1}{c!}\left(\frac{\lambda}{\mu}\right)^c \frac{\lambda^{k-c}}{\prod_{r=1}^{k-c}(c\mu + r\alpha)} p_0, \quad k > c.$$

Let us denote $\prod_{r=1}^{k-c}(c\mu + r\alpha)$ as $\xi_k$

$$\xi_k = \prod_{r=1}^{k-c}(c\mu + r\alpha) = \frac{1}{c\mu}\prod_{r=0}^{k-c}(c\mu + r\alpha) = \frac{\alpha^{k-c+1}}{c\mu}\prod_{r=0}^{k-c}\left(\frac{c\mu}{\alpha} + r\right)$$

$$= \frac{\alpha^{k-c+1}}{c\mu} \cdot \frac{\Gamma\left(\frac{c\mu}{\alpha} + k - c + 1\right)}{\Gamma\left(\frac{c\mu}{\alpha}\right)},$$

where $\Gamma(z)$ is the gamma function.

For $k > c$ we have

$$p_k = \frac{1}{c!}\left(\frac{\lambda}{\mu}\right)^c \frac{\lambda^{k-c}}{\xi_k} p_0 = \frac{1}{c!}\left(\frac{\lambda}{\mu}\right)^c \left(\frac{\lambda}{\alpha}\right)^{k-c} \frac{c\mu}{\alpha} \cdot \frac{\Gamma\left(\frac{c\mu}{\alpha}\right)}{\Gamma\left(\frac{c\mu}{\alpha} + k - c + 1\right)} p_0$$

$$= \frac{1}{c!}\left(\frac{\alpha}{\mu}\right)^c \left(\frac{\lambda}{\alpha}\right)^k \frac{c\mu}{\alpha} \cdot \frac{\Gamma\left(\frac{c\mu}{\alpha}\right)}{\Gamma\left(\frac{c\mu}{\alpha} + k - c + 1\right)} p_0.$$

The value of $p_0$ can be determined from the condition $\sum_k p_k = 1$, see Appendix A2.1. Since we do not limit the number of cruising cars, the calculation of $p_0$ involves infinite series and special functions and the resulting formula is as follows

$$p_0^{-1} = \frac{e^{\frac{\lambda}{\mu}}}{c!}\Gamma\left(c+1, \frac{\lambda}{\mu}\right) + \frac{e^{\frac{\lambda}{\alpha}}}{c!}\left(\frac{\alpha}{\mu}\right)^c \left(\frac{\lambda}{\alpha}\right)^{c\left(1-\frac{\mu}{\alpha}\right)} \gamma\left(\frac{c\mu}{\alpha} + 1, \frac{\lambda}{\alpha}\right), \tag{19}$$

where $\Gamma(x, a)$ and $\gamma(x, a)$ are upper and lower incomplete gamma functions, respectively.

In the case of the SIRO queue, we are not able to derive the closed analytical formula for the distribution of the equilibrium states' probabilities $p_k$. The problem is inherent as the $p_k$ sequence loses the form of the geometric progression. Although it is still possible to obtain an approximation of the cruising time distribution, in this case, the simulation looks to be a much simpler and more accurate alternative.

### 4.2. Characteristics of the FIFO queue model for the parking system

The major advantage of the stochastic model is in estimating the blocking probability that is, the probability that all spots are occupied. Given the arrival, departure and renege rates, the blocking probability $b$ for the FIFO model can be estimated as:

$$b = 1 - \sum_{k=0}^{c-1} p_k = 1 - p_0 \sum_{k=0}^{c-1} \frac{1}{k!}\left(\frac{\lambda}{\mu}\right)^k = 1 - \frac{e^{\frac{\lambda}{\mu}}\Gamma\left(c,\frac{\lambda}{\mu}\right)}{\Gamma(c)} p_0 = 1 - \frac{e^{c\rho}\Gamma(c,c\rho)}{\Gamma(c)} p_0 \qquad (20)$$

The value of $b$ is estimated in two ways - according to (18) and numerically (Figure 2). As can be noted, the outputs of the numeric model perfectly coincide with the theoretical estimates. Given $\rho$, the blocking probability increases with the decrease of $\alpha$.

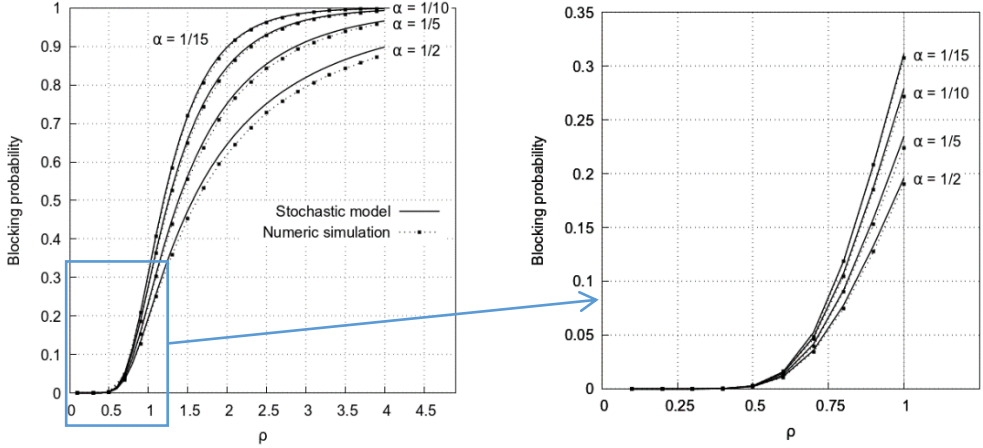

Fig. 2. Blocking probability $b$ for the stochastic FIFO model as dependent on $\rho$, for several values of $\alpha$. Parameters are $c = 20$, $\mu = \frac{1}{120}$, $\alpha = \frac{1}{2}, \frac{1}{5}, \frac{1}{10}, \frac{1}{15}$. $\lambda = \rho c \mu$.

Based on the equilibrium values of $p_k$, we can estimate the average length of the queue $Q^*$ (see appendix A2.2) and the average cruising time (time of waiting in the queue) $w_\kappa$ (see Appendix A2.3). Note that we calculate $w_\kappa$ over all cars in the system: those who found a free parking spot immediately on arrival and those cruising for some time.

$Q^*$ and $w_\kappa$ are presented in Figure 3 as dependent on $\rho$ together with the outcomes of the basic deterministic model. As can be seen, the outcomes of the deterministic model provide very good asymptotic for the averages of the stochastic FIFO model.

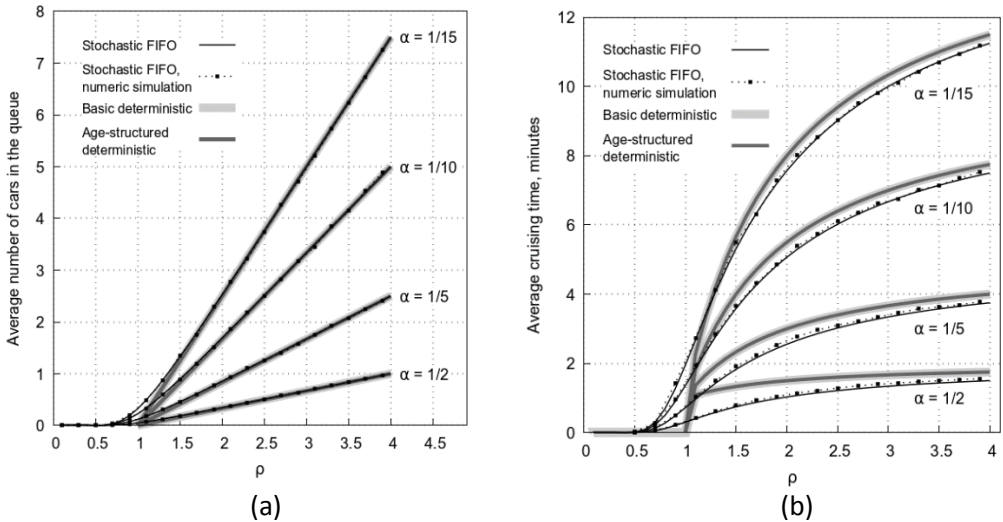

(a) (b)

Fig. 3. The average length of a queue (a), and the average cruising time $w_K$ in the queue (b) for the stochastic FIFO model. Parameters are $c = 20$, $\mu = \frac{1}{120}$, $\alpha = \frac{1}{2}, \frac{1}{5}, \frac{1}{10}, \frac{1}{15}$, $\lambda = \rho c \mu$.

The results presented in Figures 2 and 3 demonstrate a very good fit between the outcomes of the theoretical stochastic FIFO model and the numeric simulation and, also, very good asymptotic behavior of the deterministic model when it can be employed to calculate the characteristic.

Importantly, the dependency of the system's characteristics on the AD ratio $\rho$ does not change when the value of $\rho$ passes the intuitive threshold of $\rho = 1$. For $c = 20$ the blocking probability is non-zero starting from $\rho \sim 0.7$ and for the average dwell time of 2 hours ($\mu = 1/120$), the 85% rule does not guarantee parking on the lot because the blocking probability for $\rho = 0.85$ is about 0.15. At the same time, even for $\rho = 1.5$, the blocking probability for $c = 20$ is still about 0.75 that is, substantially below 1 (Figure 2). Intuitively, the blocking probability and other model characteristics should depend on $c$; let us investigate this dependency.

### 4.3. Queue model outputs' sensitivity to the value of $c$

The value of $c$ (number of parking spots in $A$) can be interpreted as the length of the search path covered by the car when cruising for parking and, intuitively, with the growth of $c$ the system becomes more deterministic. Figure 4 provides numeric estimates of this phenomenon for the reneging probability $\alpha = 1/10$. Comparing the results for $c = 20$ with those for $c = 80$, one can note that the value of $\rho$, for which the blocking probability $b$ exceeds the value of 0.1, increases from $\rho = 0.8$ to $\rho = 0.95$, while the value of $\rho$ for which the blocking probability exceeds 90% decreases from $\rho \approx 2.4$ to $\rho \approx 1.6$ (Figure 4a). At the same time, the length of the queue grows linearly proportional to the value of $c$ (Figure 4b), while given $\rho$, the average cruising time is almost insensitive to the value of $c$ (Figure 4c).

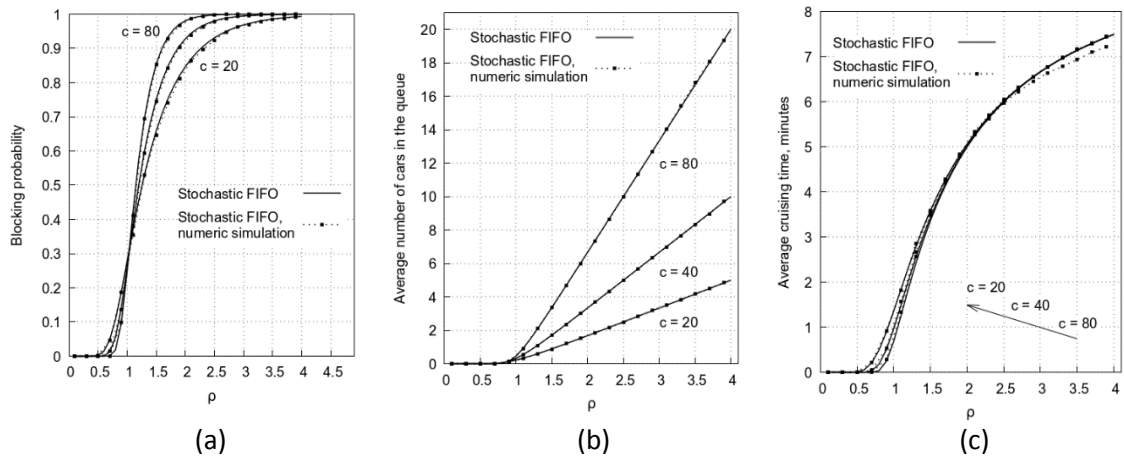

(a)  (b)  (c)

Fig. 4. Blocking probability (a), average queue length (b), and average cruising time (c) as dependent on the parking capacity $c$, for $c = 20, 40, 80$ for the FIFO discipline. $\mu = \frac{1}{120}$, $\alpha = \frac{1}{10}$, $\lambda = \rho c \mu$.

### 4.4. From FIFO to SIRO queue service discipline of the cruising cars

As mentioned, we rely upon the robustness of the average values of system performance indicators (Medhi, 2002), to relate the results obtained for the FIFO queue discipline to the SIRO discipline. Figure 5 presents the match between the results of the FIFO and SIRO models for the blocking probability, the average number of cruising cars, and the average cruising time. As can be seen, the curves for both are practically identical.

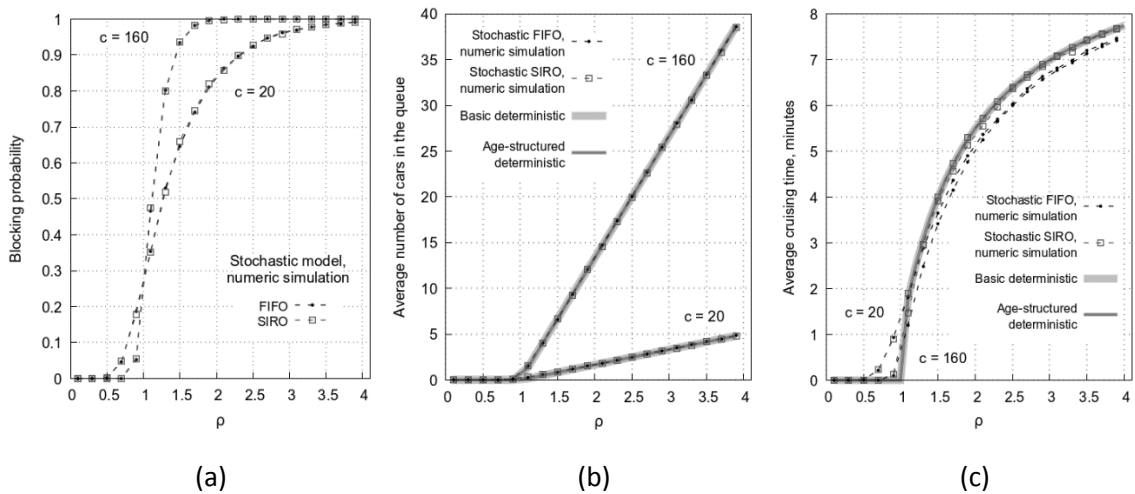

(a)          (b)          (c)

Fig. 5. Numeric simulation of the blocking probability (a), queue length (b), and average cruising time (c). FIFO and SIRO disciplines, $\mu = \frac{1}{120}$, $\alpha = \frac{1}{10}$; $c = 20$ and $c = 160$, $\lambda = \rho c \mu$.

However, as we already mentioned in section 4.1, the probability distributions of the number of cruising cars and of the cruising time for different disciplines are different. This implies that the variances of these distributions are different and so is the certainty of the drivers' parking search, which is one of the key factors determining the drivers' parking behavior. The latter difference is presented in Figure 6: for the SIRO queue, the fraction of cars that find parking in a few minutes does not decay to 0 as happens for the FIFO queue.

Fig. 6. Numeric simulation of the fraction of cars that find parking in 5 minutes for the FIFO and SIRO queue disciplines for $c = 20$ and $c = 160$, compared to the (independent of $c$) estimates obtained from

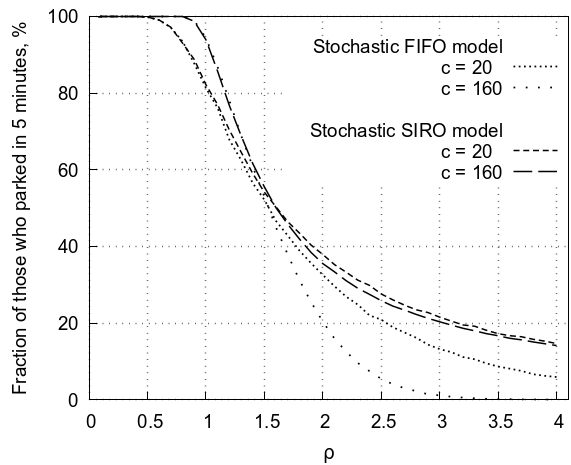

the basic deterministic model, $\mu = \frac{1}{120}, \alpha = \frac{1}{10}, \lambda = \rho c \mu$.

Our conclusion thus fully matches the theoretical view of the difference between the FIFO and SIRO models: The average characteristics of the parking dynamics are the same, while the fraction of cars that find parking after a relatively short cruising time remains substantial in the case of the SIRO queue, while for the same high values of $\rho$, this fraction for the FIFO queue is close to zero. Looking at Figure 6, we can see that for a reasonable set of parameters, this fraction is about 25% for the SIRO discipline, even for $\rho = 2$. It can also be seen in Figure 6 that the estimates of the fraction of cars that quickly find a vacant spot in the SIRO stochastic model are close to those obtained in the basic and age-structured deterministic model, while the approximation is not at the level of the deterministic approximations of the average characteristics of the parking system.

### 4.5. Non-geometric distribution of the dwell and renege times

As mentioned above, the constant departure and renege rates entail unrealistic options of either parking or reneging immediately after arrival, or after a very long parking or search time. In this section, we consider more realistic PDFs of the dwell and renege time. To compare with the case of constant $\mu$ and $\alpha$ that is, the geometric distributions of the dwell and renege times, we consider uniform distributions of these parameters with averages that are the same as for the geometric ones: the average dwell time $w_\mu = 120$ minutes and average renege time $w_\alpha = 10$ minutes. This corresponds to a constant per-minute departure rate $\mu = 1/120$ and per-minute renege rate $\alpha = 1/10$. The uniform distributions that we consider in this section are

- Uniform on the interval [0.5, 3.5] hours distribution of the dwell time
- Uniform on the interval [0, 20] minutes distribution of the renege time

Figure 7 presents these and geometric distributions for the visual comparison.

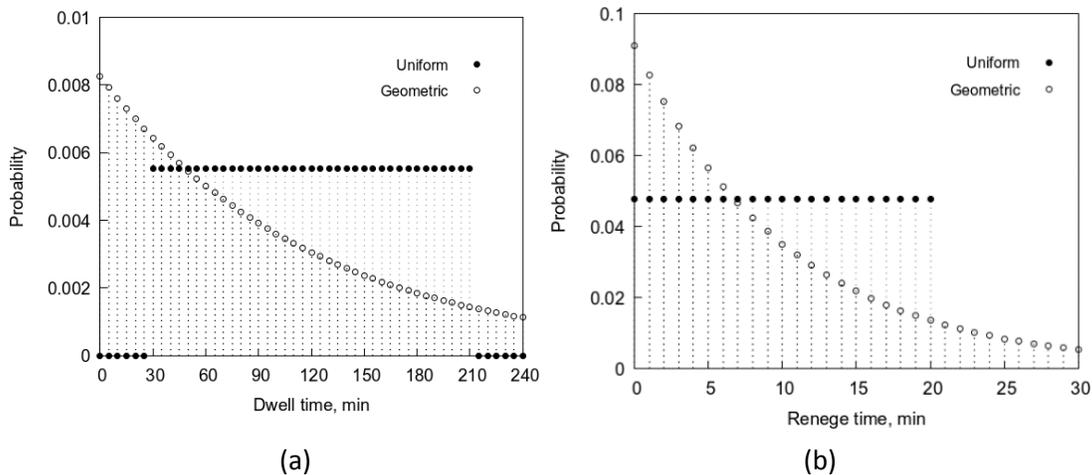

(a)                                     (b)

Figure 7. Uniform and geometric distributions of the dwell time and renege time with the common averages (a) dwell time, $w_\mu = 120$; (b) renege time, $w_\alpha = 10$.

Representing the AD ratio $\rho = \lambda/\mu c$ via the average dwell time $w_\mu$, $\rho = \lambda w_\mu/c$, we can express the probability $\varepsilon$ that the cruising car will find parking (see (8)), and expression for the fraction $\phi$ of cars that will continue cruising in the case of $\lambda w_\mu > c$ via average dwell and renege times:

$$\varepsilon = \frac{c\mu\alpha}{\lambda - c\mu + c\mu\alpha} = \frac{c}{\lambda w_\mu w_\alpha - c w_\alpha + c}, \qquad (21)$$

$$\phi = (1-\varepsilon)(1-\alpha) = \frac{(\lambda w_\mu - c)(w_\alpha - 1)}{\lambda w_\mu w_\alpha - c w_\alpha + c}, \qquad (22)$$

The equilibrium characteristics of the deterministic model (3) become thus:

$$Q^* = \frac{\lambda - c\mu}{\alpha} = (\lambda w_\mu - c)\frac{w_\alpha}{w_\mu}, \qquad (23)$$

$$f^* = \frac{c\mu}{\lambda} = \frac{c}{\lambda w_\mu}, \qquad (24)$$

$$w_K = \frac{(1-\varepsilon)(1-\alpha)}{\varepsilon + \alpha - \varepsilon\alpha} = \frac{(\lambda w_\mu - c)(w_\alpha - 1)}{\lambda w_\mu}, \qquad (25)$$

While the fraction $\psi_\tau$ of cars arriving in the system and finding parking in $\tau$ or less time steps is:

$$\psi_\tau = \frac{c(1-\phi^{\tau+1})/(1-\phi)}{\lambda w_\mu w_\alpha - c w_\alpha + c} = \frac{c(1-\phi^{\tau+1})}{\lambda w_\mu} = f^*(1-\phi^{\tau+1}). \qquad (26)$$

Figure 8 compares model outputs for the geometric and uniform distributions of the dwell and renege times, for $c = 20$. When possible, the deterministic outcomes are superimposed.

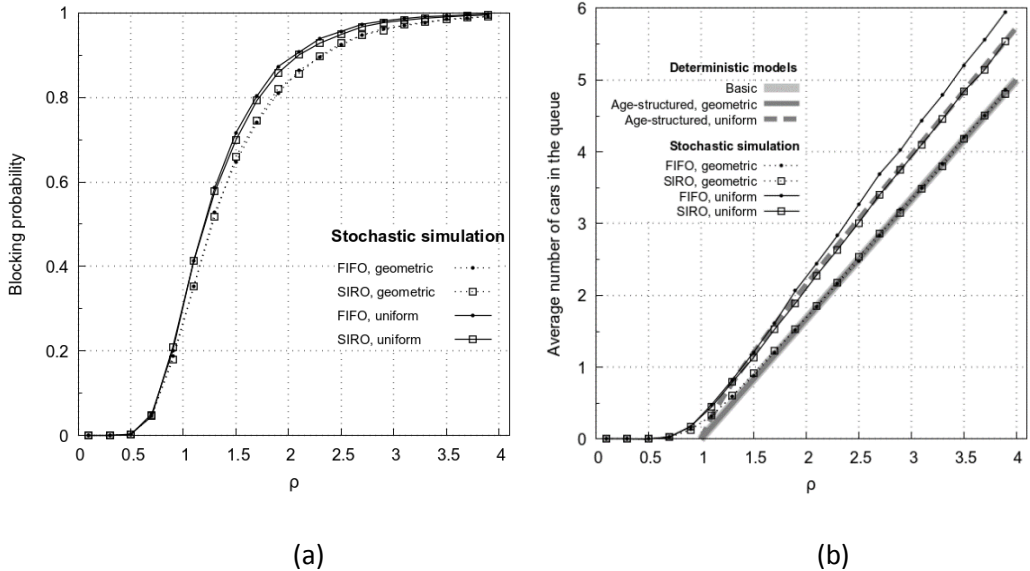

(a)            (b)

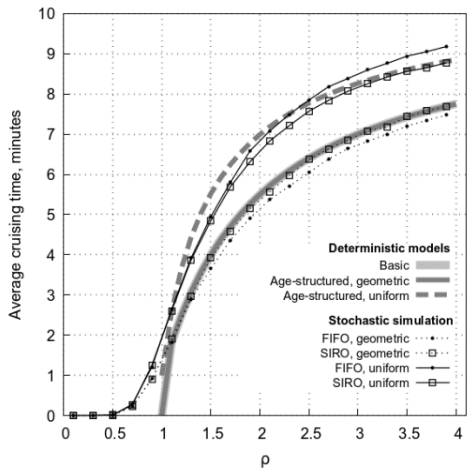

(c)

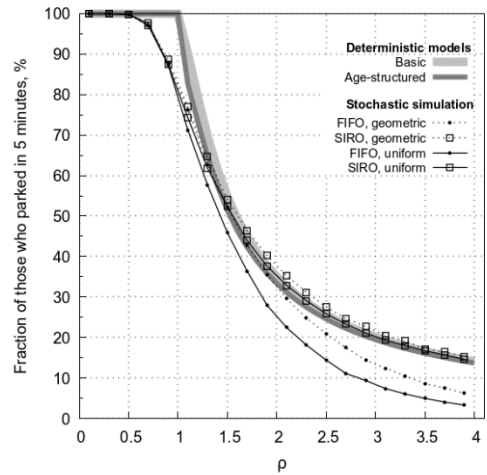

(d)

Fig. 8 Numeric simulation of the FIFO and SIRO queue disciplines with the outcomes of two deterministic models superimposed, $c = 20$. (a) blocking probability; (b) average queue length; (c) average cruising time; (d) fraction of drivers who park in less than 5 minutes. $\mu = 1/120$ for the geometric PDF and $w_\mu = 120$ for the $U(30, 210)$ distribution of the dwell time. $\alpha = 1/10$ for the geometric PDF and $w_\alpha = 10$ for the $U(0, 20)$ distribution of the renege time.

Figure 9 presents the same outputs for $c = 160$.

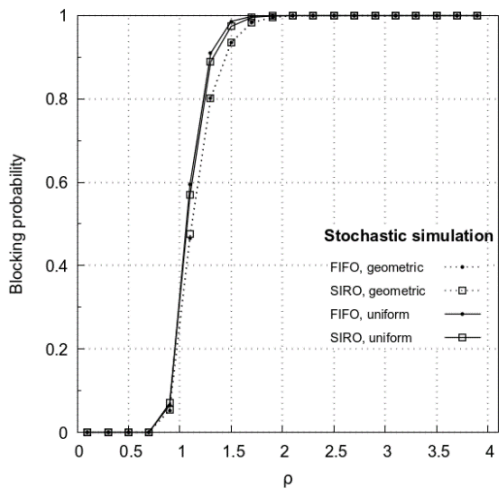

(a)

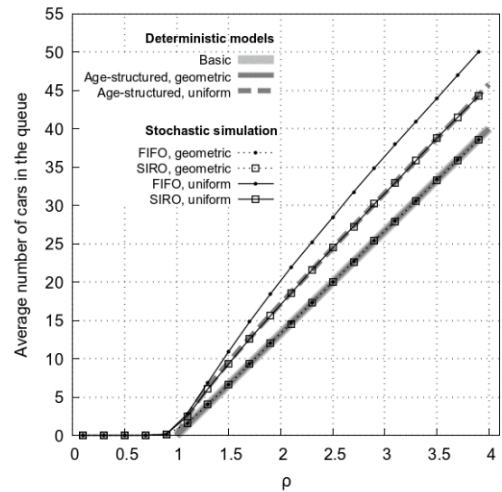

(b)

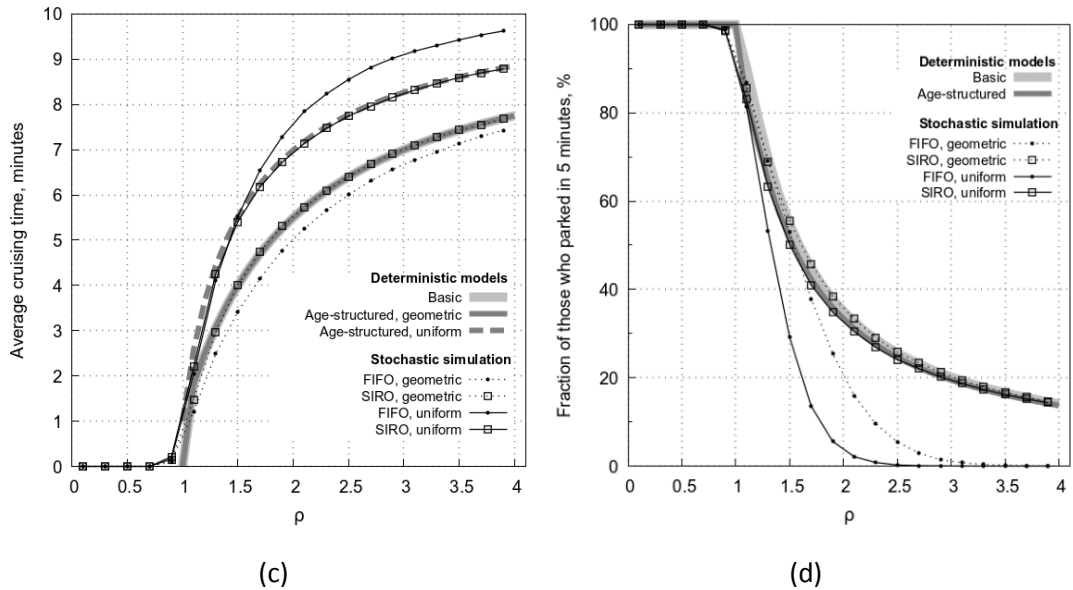

(c)                                   (d)

Fig. 9. Numeric simulation of the FIFO and SIRO disciplines with the outcomes of two deterministic models superimposed, $c = 160$. (a) blocking probability; (b) average queue length; (c) average cruising time; (d) fraction of drivers who park in less than 5 minutes. $\mu = 1/120$ for the geometric PDF and $w_\mu = 120$ for the $U(30, 210)$ distribution of the dwell time. $\alpha = 1/10$ for the geometric PDF and $w_\alpha = 10$ for the $U(0, 20)$ distribution of the renege time.

As we can see, the outputs of the SIRO models with geometric and uniform distributions of the dwell and renege times are close or very close to each other and also to the outputs of the deterministic models. Importantly, irrespective of the shape of the distributions, the fraction of cars that park for less than 5 minutes remains at the level of 20% up to very high values for the AD ratio $\rho$. The major difference between the FIFO and SIRO queue models that is characteristic for both types of dwell and renege time distributions is the fraction of cars that find parking within a few minutes.

## 5. Discussion

The goal of this paper is to build a simple theoretical background for the analysis of parking dynamics. Its outcomes – number of cruising cars, cruising time, fraction of cars that succeed to find a parking spot, and the probability to park in less than $\tau$ minutes (23) – (26) can be used for estimating the parameters of the parking system depending on its capacity, dynamics of arrivals, departures and drivers' willingness to cruise. Besides importance of these results for studying on-street parking dynamics, our findings may be used as a component of the more complex parking models that account for the competition between on-street and off-street parking, including competition between parking lots, drivers' reactions to on- and off-street pricing policies, and illegal parking. Such models can be utilized for more effective parking control and establishing appropriate on-street parking prices. The obtained simple approximation of the on-street parking dynamics characteristics can be further exploited in the general traffic model where drivers may react to lengthy cruising times with long-term behavioral changes, like the drift of departure or arrival times, or a modal shift. Below, we sketch the ways of incorporating these changes into the proposed models and thereby bridging our theoretical models to reality.

First, some of the real-world phenomena do not demand change to the models. For example, the dependence of the major model parameters $\lambda$, $\mu$, or $\alpha$ on (1) the hour of day or (2) the accumulated cruising time. These dependencies, if available, can be directly employed in the age-structured model (11) – (14). Let us briefly consider three other qualitative phenomena demanding a closer view: (1) spatial heterogeneity of the parking demand and supply; (2) long-term consequences of lengthy cruising times; and (3) the possible effect of the *occupation lag* – the time lag between the moment a spot is vacated and the moment one of the cruising cars occupies it.

### 5.1. The unbounded heterogeneous parking space

Our models consider the parking search in the bounded homogeneous area $A$ that contains $c$ parking spots, and the goal of a driver is to park somewhere in $A$. In reality, drivers search for parking in heterogeneous and unbounded urban space, and their destinations are *buildings*. Let us take a step towards reality and assume that drivers whose destination is a building $B$ traverse while cruising, $B$'s *on-street parking search neighborhood* $U(B)$ that contains $c_B$ parking spots $s_i, i = 1, ..., c_B$ (Figure 10a). In reality, the shape and size of the search neighborhood depends on the street topology, characteristic of the roads, curb parking permissions. To simplify, let us also assume that the probability of visiting each spot in $U(B)$ is the same, $1/c_B$ and ignore the possible dependence of this probability on the distance between the destination $B$ and the spot, street topology, parking permissions, and other factors that can affect drivers' cruising behavior. Let us denote the demand for parking at $B$ as $\omega_B$ and simplify the problem even more by assuming that the average dwell and renege times $w_\mu$ and $w_\alpha$ are the same for all drivers.

The next step is distributing the demand $\omega_B$ over the $U(B)$: If every spot in $U(B)$ is in equal demand by the drivers whose destination is $B$, then for each $s_i \in U(B)$ the demand is $\omega_B/c_B$. Importantly, every spot $s_i$ can serve other cars – those whose destination is any building $B_j$, such that $s_i \in U(B_j)$. As a result, the overall demand $\lambda_{s_i}$ for each parking spot $s_i$ is equal to

$$\lambda_{s_i} = \sum_{over\ all\ B_j\ for\ which\ s_i \in U(B_j)} \omega_{B_j}/c_{B_j} \qquad (27)$$

Finally, the parking conditions for the drivers whose destination is $B$ will be characterized by the arrival rate $\lambda_{U(B)}$ equal to the total demand of all spots in $U(B)$:

$$\lambda_{U(B)} = \sum_{s_i \in U(B)} \lambda_{s_i} \qquad (28)$$

Figure 10(b) presents the exemplar map of $\lambda_{U(B)}$, by buildings, constructed based on the buildings' demand as presented in Figure 10(a), for the search neighborhood radius chosen, for simplicity, as an area of a radius 300m (Levy, Benenson, 2015; Dowling, 2020) with a building in a center. This map is the basis for estimating the major parameters of the parking system, like the average number of cars cruising in the $U(B)$, or the average cruising time for cars arriving to $B$.

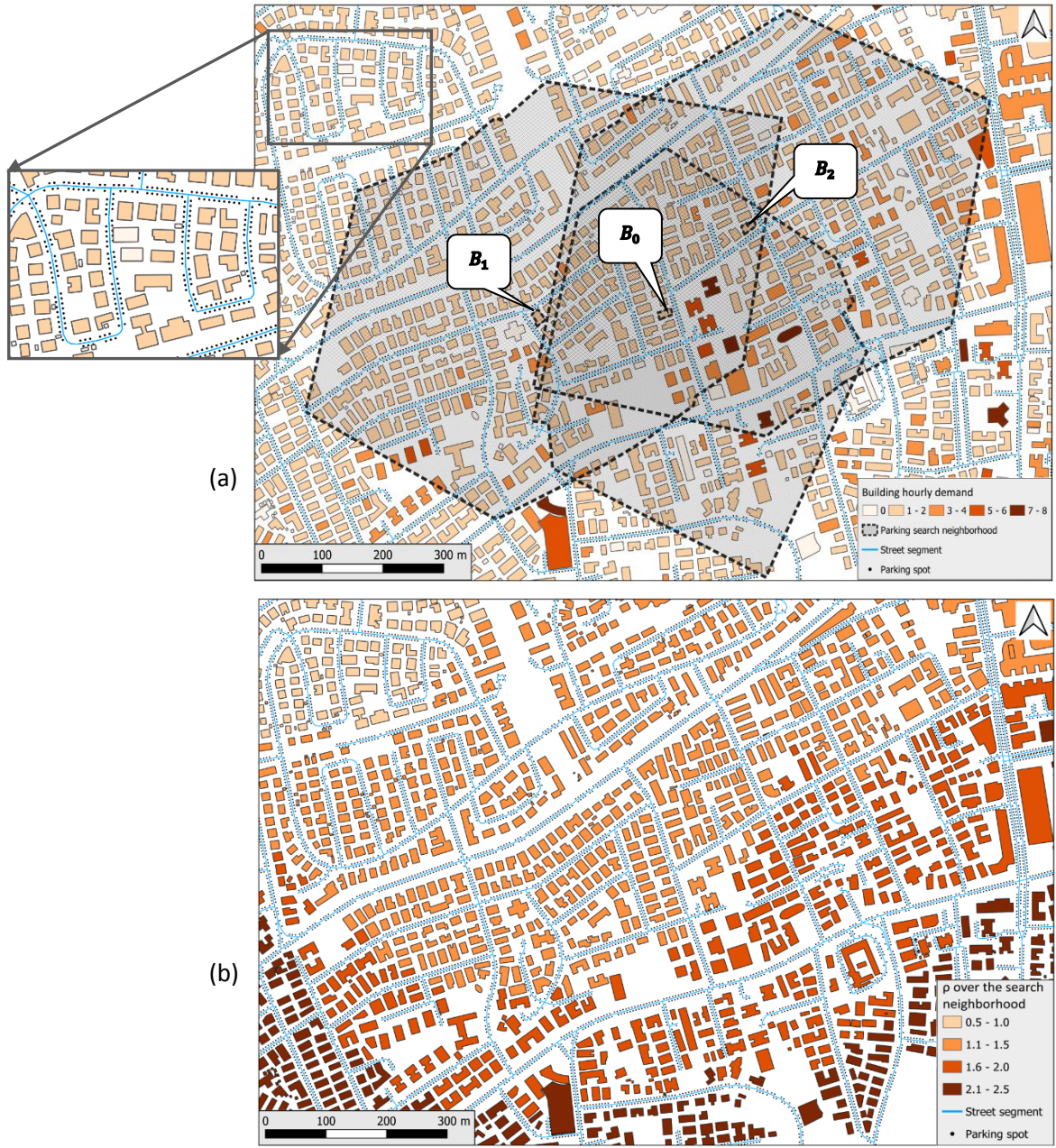

Figure 10: (a) Parking demand by buildings and schematic representation of the overlapping on-street parking search neighborhoods for three nearby buildings, $B_0$, $B_1$ and $B_2$. Curb parking spots are at a distance of 5 m between each other, see inset; (b) map of the parking search conditions expressed via the value of $\rho$ for the drivers whose destination is in each of the area's buildings.

## 5.2. Dependence of drivers' behavior on accumulated search time

The longer the cruising time, the stronger the driver's frustration, which increases the driver's willingness to cancel this search and never cruise here again. This desire can be formalized in two ways: In the short term, we can assume that driver's probability to renege $\alpha$ increases when cruising longer. This dependency can be explicitly reflected by the age-structured model (11) – (14). In the long term, when visiting the area the next time, a driver can decide to go directly to the off-street parking lot, use public transport, or just cancel the visit, and these phenomena can be interpreted as the dependence of "tomorrow's" arrival rate on "today's" cruising time.

The interpretation of a driver's long-term reaction can be that the drivers who have searched for too long at day $T$ will not participate in the on-street parking search at day $T + 1$. Formally, let us assume that every day the parking area $A$ attracts a constant number $\lambda_n$ of new drivers, who arrive there in the hope that their parking search will not be too long, while drivers who cruise for too long do not return the next day. Let us interpret "cruising too long" as longer than the average cruising time $1/\alpha$ corresponding to the "standard" drivers' habits.

The dynamics of arrivals in these circumstances may be described as

$$\lambda(T+1) = \lambda(T)(1 - df_\tau(T)) + \lambda_n \qquad (29)$$

where $f_\tau(T)$ is the fraction of cars that cruise for longer than $\tau$ at day $T$. Note that $f_\tau(T)$ is defined by the value of $\rho$ at day $T$ and is calculated according to (10).

To illustrate the effect of the possible dependency (29), let us consider the system with the parameters $c = 160, \mu = \frac{1}{120}, \lambda_n = 0.5c\mu$ and varying $d$, and simulate the dynamics of arrivals, in days, for the value of $\alpha = 1/10$ that we use in all our examples. Figure 11 presents the cobweb plot (Hoffman and Frenkel, 2001) for the discrete-time recurrent equation (29) in terms of $\rho$, for different values of $d$:

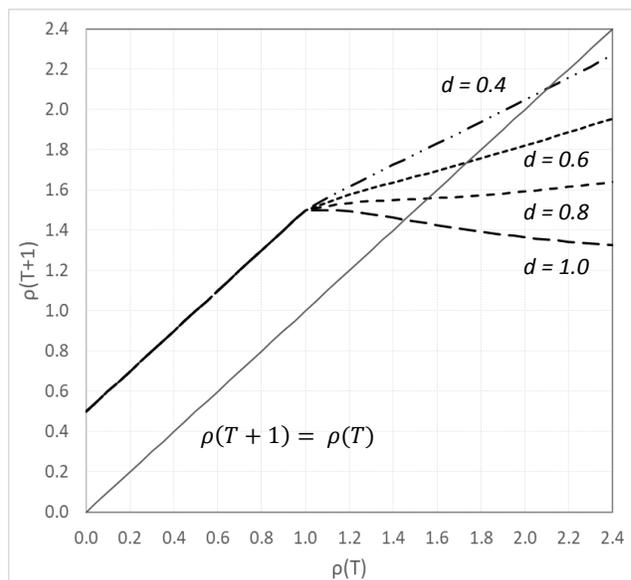

Figure 11. A cobweb plot for the $\rho(T+1)$ as dependent on $\rho(T)$, for $c = 160, \mu = \frac{1}{120}, \alpha = \frac{1}{10}, \lambda_n = 0.5c\mu$ and $d = 1.0, 0.8, 0.6,$ and $0.4$.

As can be seen in Figure 11, for all values of $d$, system (29) converges to the stable equilibrium that increases with the decrease in $d$ that is, with the decrease in the fraction of drivers who react to too long cruising. For the investigated values of parameters, the equilibrium value of $\rho$ increases from $\rho = 1.32$ for $d = 1$ to $\rho = 2.06$ for $d = 0.4$.

### 5.3. The occupation lag

All our models include the unarticulated assumption that when the spot is vacated, one of the cruising cars occupies it during the same time step of one minute. This may be indeed close to reality when $\rho \gg 1$: Considering the equilibrium number of cruising cars $Q^*$ relative to the number of parking spaces $c$ for the values of $w_\alpha = 10\ min$ and $w_\mu = 120\ min$ that we consider typical, we can see that, according to (4), the ratio of the number of cruising cars to the number of spots in the search area is equal to $\frac{(\rho-1)\mu}{\alpha}$. The numeric value of this ratio is about ~8.5% for $\rho = 2$, 17% for $\rho = 3$, and 25.5% for $\rho = 4$ similar to Shoup's (2011) estimates of the percentages of cruising drivers among the general traffic over highly demanded parking areas. A high density of cruising cars for the large values of $\rho \gg 1$ guarantees that the *occupation lag* will be short: For example, if the cruising cars are spread over the search area $A$ uniformly, then, for $\rho = 3$, the distance between them is $1/0.17 \approx 6$ spots only. For the average cruising speed of 12-18 km/h (Fulman & Benenson, 2021), the occupation lag in this case will be about 10 seconds.

The occupation lag is defined by the drivers' cruising behavior and the number of cruising cars; the lower the number of cars, the higher the lag and the lower the share $\theta$ of the vacant spots that will be occupied during the same time step. If the dependency of $\theta$ on the queue length is known, it can be incorporated into the aggregate model (4) as $\theta = \theta(Q(t)) \leq 1, d\theta/dQ < 0$.

To reflect the occupation lag, equations (4) should be modified as follows:

$$P(t+1) = (1-\mu)P(t) + \theta\left(c - (1-\mu)P(t)\right) \quad (30)$$

$$Q(t+1) = (1-\alpha)Q(t) + \lambda - \theta\left(c - (1-\mu)P(t)\right)$$

$$\theta = \theta(Q(t))$$

The equilibrium value of the number of occupied spots for (30) is lower than $c$:

$$P^* = c - \frac{(1-\theta)c\mu}{\mu+\theta-\mu\theta} \quad (31),$$

where $\frac{(1-\theta)c\mu}{\mu+\theta-\mu\theta}$ is the non-zero equilibrium number of vacancies.

The equilibrium length of the queue is larger than obtained for the zero occupation lag:

$$Q^* = \frac{\lambda - c\mu}{\alpha} + \frac{(1-\theta)c\mu^2}{\alpha(\mu+\theta-\theta\mu)} \qquad (32).$$

Further study of the effect of the occupation lag demands knowledge of the dependency of $\theta$ on $Q(t)$ and we limit ourselves to the numeric examples of the equilibrium for $\theta = 0.5$ only:

$$P^* = c - \frac{0.5c\mu}{\mu + 0.5 - 0.5\mu} = c - \frac{c\mu}{\mu+1}$$

$$Q^* = \frac{\lambda - c\mu}{\alpha} + \frac{0.5c\mu}{\alpha(0.5\mu + 0.5)} = \frac{\lambda - c\mu}{\alpha} + \frac{c\mu}{\alpha(\mu+1)}$$

For the values of $\mu = 1/120$ and $\alpha = 1/10$ that we employ in this paper, this equilibrium is very close to that for the zero occupation lag.

## 6. Conclusions

Our analysis of deterministic and stochastic parking models reveals several basic facts: Given the parking area $A$ with $c$ spots, arrival rate $\lambda$, and departure rate $\mu$ or average dwell time $w_\mu$, the state of the parking system is defined by the Arrivals-to-Departures (AD) ratio $\rho = \lambda/(c\mu) = \lambda w_\mu/c$ that can be considered as a formal expression for the generally used "demand-to-supply" ratio. With an increase of $\rho$, the state of parking in $A$ evolves from "immediate parking" to "cruising for parking," and there is no crisp boundary between these states. For a small parking area $A$, the probability to be fully occupied is non-zero starting from $\rho \sim 0.7$, while for a large $A$, the probability that $A$ will be fully occupied (the blocking probability) is non-zero when $\rho$ exceeds 0.9. Shoup's 85% threshold (Shoup 2011) is, thus, a wonderful compromise. The blocking probability grows non-linearly with the growth of $\rho$ and, for the likely values of $c$, becomes close to 1 when $\rho$ exceeds $2 - 2.5$. In these circumstances, the occupation of a vacated spot by one of the cruising cars is almost immediate (during the same time step in the model).

Our results are based on the assumption that the driver's chance to find a vacant spot does not depend on her/his accumulated cruising time. This assumption is explicitly employed in both deterministic models and interpreted as the Service-In-Random-Order (SIRO) discipline in the probabilistic model of parking. However, the average values of the major system indicators, like the average number of cruising cars or average cruising time, are practically the same for the SIRO and FIFO (First-In-First-Out) queueing disciplines. Importantly, the basic and age-structured deterministic models provide a very good approximation of these averages, and simple formulas $Q^* = (\lambda w_\mu - c)\frac{w_\alpha}{w_\mu}$ for the average number of cruising cars $Q^*$, and $w_\kappa = \frac{(1-\alpha)(\lambda w_\mu - c)w_\alpha}{\alpha(\lambda w_\mu - c)w_\alpha + c}$ for the average cruising time $w_\kappa$, work very well. To recall, $\lambda$ is the arrival rate, $w_\mu$ is the average dwell time of the parking cars and $w_\alpha$ is the average renege time of the cruising cars. Using the AD ratio $\rho$, the formulas for $Q^*$ and $w_\kappa$ can be presented as $Q^* = c(\rho - 1)\frac{w_\alpha}{w_\mu}$ and $w_\kappa = (w_\alpha - 1)\frac{\rho-1}{\rho}$, that is, with the growth of $\rho$, the number of the cruising cars grows linearly, while the average cruising time decreases hyperbolically.

Our theoretical analysis of the probabilistic models is based on the assumption of constant departure and renege rates. For the constant rates, the basic deterministic approximation works wonderfully not only for the average system indicators but also for the characteristics of the search time distribution and the fraction of cars that cruise less than a threshold time (5 minutes in our numeric examples). In a more realistic case of uniform distributions of the dwell and renege time, the approximation provided by the age-structured deterministic model is better than that of the basic one. However, the two model's outcomes differ by a level of 10-20%, at most. Importantly, irrespective of the shape of the distributions, the fraction of cars that park for less than 5 minutes remains at the level of 20% up to very high values for the AD ratio $\rho$.

It is important to note that our framework does not include many of the complexities of the parking reality. Real-world drivers are heterogeneous regarding their aversion to the risk of parking failure and may employ search tactics that provide a real or illusory advantage in competition with other drivers (Cassady, Kobza, 1998; Krapivsky, Redner 2019). Some drivers (including the authors of this paper) prefer to park at the first vacant spot they encounter within a 5- or even 10-minute walking distance from the destination; others double park at a vantage point close to the destination where a large number of occupied on-street spots are visible and wait there for a departing car or for drivers walking to their cars. Another popular tactic is cruising close to the destination looking for various signs of imminent car departure (Rubak, Juieng, 1997; Kokolaki, Stavrakakis, 2014; Pawlowsky, 2016; Rosni et al, 2019). In this paper, we investigate parking dynamics as defined by the basic parameters of arrival, departure, and search reneging.

To conclude, we provide simple analytical expressions for the major parameters of the parking system and confirm their robustness. At the same time, our model framework is non-spatial. Estimating the distance between the parking spot and the destination requires an explicit account of the *spatial* behavior of the parking drivers that we ignored in this paper. The results of the investigation of the abstract but spatial parking model will be presented in a future paper.

## 7. Acknowledgments

This research was supported by the ISRAEL SCIENCE FOUNDATION (grant No. 1160/18) "Tessellation of urban parking prices". AO is partially supported by the Israeli Ministry of Absorption.

*Appendix 1*

**Notation**

**A1.1. Parking area**

$A$ – the area where the driver searches for parking

$c$ – number of parking spots in $A$

**A1.2 Dynamic models**

$t$ – time in discrete units (time steps)

**A1.2.1. Constant dwell and renege rates**

$P(t)$ – the total number of cars that parked in $A$ at $t$

$\lambda$ – arrival rate, the average number of cars arriving per time step

$\mu$ – departure rate, the probability to leave per time step

$\alpha$ – renege rate for cruising cars, the probability to leave the queue per time step

$\rho = \lambda/(c\mu)$ – arrivals to departure per spot ratio

**A1.2.2. Arbitrarily distribution of dwell and renege rates (age-structured model)**

$s_{max}$ – maximal dwell time

$v_{max}$ – maximal renege time

$w_\mu$ – average dwell time

$w_\alpha$ – average renege time

$P^s(t)$ – number of cars who parked at $A$ at a time moment $t - s$

$\bar{P}(t) = \left(P^0(t), \ldots, P^s(t), \ldots, P^{s_{max}}(t)\right)^T$ – vector representing the number of cars that parked after cruising for t time steps

$Q^v(t)$ – number of cars who entered $A$ at a time moment $t - v$

$\bar{Q}(t) = \left(Q^0(t), \ldots, Q^v(t), \ldots, Q^{v_{max}}(t)\right)^T$ - representing the number of cars that reneged after cruising for t time steps

**A1.2.3. Stochastic model**

$p_n(t)$ – the probability that there are $n$ cars either parked or waiting for a vacant spot

$b$ – blocking probability, the probability that all $c$ spots in $A$ are occupied

$\gamma_v$ – the probability that the driver reneges after exactly $v$ time units of cruising

$U(a, b)$ – uniform distribution on *[a, b]*

### A1.2.4. Equilibrium state of the system

$w_\kappa$ – average equilibrium cruising time

$P^*$ - equilibrium number of parked cars

$Q^*$ - equilibrium length of the queue of cruising cars

$f^*$ – equilibrium fraction of cars, among those that arrived at the system, that succeeded to park

$\varepsilon$ – equilibrium probability that the cruising car will find parking

$\phi$ – equilibrium probability that the car will continue cruising

$\psi_\tau$ – equilibrium fraction of cars that would find parking in τ time steps or less

$\delta_s$ – equilibrium fraction of cars who park for exactly $s$ time steps

*Appendix 2*

## Characteristics of the stochastic queue model with reneging

We are not aware of any publication that presents together three critical for parking application characteristics of the FIFO M/M/c model with reneging - blocking probability, queue length, and waiting time distribution. We also employ them for verifying the numeric modeling results. Below we compile all necessary theoretical results, following the unified notation that was used throughout the paper. Sections A2.1 and A2.2 complete our derivation of the probabilities of the system's states that is started in section 4.1. Section A2.3 adapts the waiting time distribution of Ancker and Cafarian (1963) to our parking application, and section A2.4 presents several formulas that we employed in this appendix for the reader interested in the mathematical details.

### A2.1 Empty system probability $p_0$

From the identity $\sum_k p_k = 1$ it follows that

$$p_0^{-1} = \sum_{k=0}^{\infty} \frac{p_k}{p_0} = \underbrace{\sum_{k=0}^{c} \frac{1}{k!}\left(\frac{\lambda}{\mu}\right)^k}_{S_1} + \frac{1}{c!}\left(\frac{\alpha}{\mu}\right)^c \frac{c\mu}{\alpha} \Gamma\left(\frac{c\mu}{\alpha}\right) \underbrace{\sum_{k=c+1}^{\infty} \frac{\left(\frac{\lambda}{\alpha}\right)^k}{\Gamma\left(\frac{c\mu}{\alpha}+k-c+1\right)}}_{S_2}.$$

The first sum can be easily expressed through the upper incomplete gamma function:

$$S_1 = \frac{e^{\frac{\lambda}{\mu}}}{c!} \Gamma\left(c+1, \frac{\lambda}{\mu}\right).$$

Second term in $p_0^{-1}$ requires more work:

$$S_2 = \sum_{k=c+1}^{\infty} \frac{\left(\frac{\lambda}{\alpha}\right)^k}{\Gamma\left(\frac{c\mu}{\alpha}+k-c+1\right)} = \sum_{r=0}^{\infty} \frac{\left(\frac{\lambda}{\alpha}\right)^{r+c+1}}{\Gamma\left(\frac{c\mu}{\alpha}+2+r\right)} = \left(\frac{\lambda}{\alpha}\right)^{c+1} \sum_{r=0}^{\infty} \frac{\left(\frac{\lambda}{\alpha}\right)^r}{\Gamma\left(\frac{c\mu}{\alpha}+2+r\right)}$$

$$= \frac{e^{\frac{\lambda}{\alpha}}\left(\frac{\lambda}{\alpha}\right)^{c\left(1-\frac{\mu}{\alpha}\right)}}{\frac{c\mu}{\alpha}} \cdot \frac{\gamma\left(\frac{c\mu}{\alpha}+1,\frac{\lambda}{\alpha}\right)}{\Gamma\left(\frac{c\mu}{\alpha}\right)},$$

where $\gamma(x, a)$ is a lower incomplete gamma function. Summing up, we have finally

$$p_0^{-1} = \frac{e^{\frac{\lambda}{\mu}}}{c!}\Gamma\left(c+1,\frac{\lambda}{\mu}\right) + \frac{1}{c!}\left(\frac{\alpha}{\mu}\right)^c \frac{c\mu}{\alpha}\Gamma\left(\frac{c\mu}{\alpha}\right) \cdot \frac{e^{\frac{\lambda}{\alpha}}\left(\frac{\lambda}{\alpha}\right)^{c\left(1-\frac{\mu}{\alpha}\right)}}{\frac{c\mu}{\alpha}} \cdot \frac{\gamma\left(\frac{c\mu}{\alpha}+1,\frac{\lambda}{\alpha}\right)}{\Gamma\left(\frac{c\mu}{\alpha}\right)}$$

$$= \frac{e^{\frac{\lambda}{\mu}}}{c!}\Gamma\left(c+1,\frac{\lambda}{\mu}\right) + \frac{e^{\frac{\lambda}{\alpha}}}{c!}\left(\frac{\alpha}{\mu}\right)^c \left(\frac{\lambda}{\alpha}\right)^{c\left(1-\frac{\mu}{\alpha}\right)} \gamma\left(\frac{c\mu}{\alpha}+1,\frac{\lambda}{\alpha}\right).$$

### A2.2 Average queue length $Q^*$

Having $p_0$, we can calculate the average number $Q^*$ of cars in the queue (that is cruising in search of the free parking spot):

$$\frac{Q^*}{p_0} = \sum_{k=c}^{\infty}(k-c)\frac{p_k}{p_0} = \frac{1}{c!}\left(\frac{\alpha}{\mu}\right)^c \frac{c\mu}{\alpha} \Gamma\left(\frac{c\mu}{\alpha}\right) \underbrace{\sum_{k=c}^{\infty} \frac{(k-c)\left(\frac{\lambda}{\alpha}\right)^k}{\Gamma\left(\frac{c\mu}{\alpha}+k-c+1\right)}}_{A}.$$

The term $A$ can be calculated in terms of incomplete gamma function:

$$A = \sum_{k=c}^{\infty} \frac{(k-c)\left(\frac{\lambda}{\alpha}\right)^k}{\Gamma\left(\frac{c\mu}{\alpha}+k-c+1\right)} = \left(\frac{\lambda}{\alpha}\right)^c \sum_{r=0}^{\infty} \frac{r\left(\frac{\lambda}{\alpha}\right)^r}{\Gamma\left(\frac{c\mu}{\alpha}+r+1\right)} =$$

$$= \left(\frac{\lambda}{\alpha}\right)^c \cdot \frac{\left(\frac{c\mu}{\alpha}+1\right)\left(\frac{\lambda}{\alpha}\right)^{\frac{c\mu}{\alpha}}\left[\left(\frac{\lambda}{\alpha}\right)^{\frac{c\mu}{\alpha}+1} + e^{\frac{\lambda}{\alpha}}\left(\frac{\lambda}{\alpha}-\frac{c\mu}{\alpha}\right)\gamma\left(\frac{c\mu}{\alpha}+1,\frac{\lambda}{\alpha}\right)\right]}{\Gamma\left(\frac{c\mu}{\alpha}+2\right)}$$

$$= \left(\frac{\lambda}{\alpha}\right)^c \cdot \frac{\frac{\lambda}{\alpha}\left(\frac{c\mu}{\alpha}+1\right) + e^{\frac{\lambda}{\alpha}}\cdot\left(\frac{\lambda}{\alpha}\right)^{-\frac{c\mu}{\alpha}}\left(\frac{c\mu}{\alpha}+1\right)\left(\frac{\lambda}{\alpha}-\frac{c\mu}{\alpha}\right)\gamma\left(\frac{c\mu}{\alpha}+1,\frac{\lambda}{\alpha}\right)}{\left(\frac{c\mu}{\alpha}+1\right)\Gamma\left(\frac{c\mu}{\alpha}+1\right)}$$

$$= \frac{\left(\frac{\lambda}{\alpha}\right)^{c+1} + e^{\frac{\lambda}{\alpha}}\cdot\left(\frac{\lambda}{\alpha}\right)^{c\left(1-\frac{\mu}{\alpha}\right)}\left(\frac{\lambda}{\alpha}-\frac{c\mu}{\alpha}\right)\gamma\left(\frac{c\mu}{\alpha}+1,\frac{\lambda}{\alpha}\right)}{\frac{c\mu}{\alpha}\Gamma\left(\frac{c\mu}{\alpha}\right)}$$

Now we can find $Q^*$:

$$\frac{Q^*}{p_0} = \frac{1}{c!}\left(\frac{\alpha}{\mu}\right)^c \left[\left(\frac{\lambda}{\alpha}\right)^{c+1} + e^{\frac{\lambda}{\alpha}}\cdot\left(\frac{\lambda}{\alpha}\right)^{c\left(1-\frac{\mu}{\alpha}\right)}\left(\frac{\lambda}{\alpha}-\frac{c\mu}{\alpha}\right)\gamma\left(\frac{c\mu}{\alpha}+1,\frac{\lambda}{\alpha}\right)\right].$$

### A2.3 The waiting time distribution and its average value $w_\kappa$

In calculating the waiting time distribution, we follow Ancker & Cafarian (1963), specifically, the case of the system with an unbounded waiting queue and identical parking spots. Here we present the main definitions and the principal results without going into technical details of their derivation.

$$c_0 = 1;\ c_r = \frac{c}{r}\left(\frac{\mu}{\lambda}\right)^r, r = 1,2,\dots,c;$$

The probability of a full parking system with an empty queue is given by $p_c = \frac{1}{F+G}$, where

$$F = \sum_{j=1}^{c} c_j j! = c! \left(\frac{\mu}{\lambda}\right)^c e^{\frac{\lambda}{\mu}} \left[1 - I\left(\frac{\lambda}{\mu\sqrt{c}}, c - 1\right)\right],$$

$$G = \Gamma\left(\frac{\mu}{\alpha} + 1\right) \sum_{j=0}^{\infty} \frac{\left(\frac{\lambda}{\alpha}\right)^j}{\Gamma\left(\frac{\mu}{\alpha} + 1 + j\right)} = \frac{\Gamma\left(\frac{\mu}{\alpha} + 1\right)}{e^{-\frac{\lambda}{\mu}}\left(\frac{\lambda}{\mu}\right)^{\frac{\mu}{\alpha}}} I\left(\frac{\lambda}{\sqrt{\alpha\mu}}, \frac{\mu}{\alpha} - 1\right),$$

$$I(u, p) = \frac{\int_0^{u\sqrt{p+1}} e^{-v} v^p dv}{\Gamma(p + 1)} \text{ — incomplete gamma function ratio.}$$

Probability of the event $A$ that an arbitrary cruising car acquires parking service:

$$P[A] = \frac{\mu}{\alpha} \cdot \frac{G - 1}{G}$$

In regard to the waiting time distribution, it is important to distinguish between the time spent in the queue (that is cruising and searching for a free spot) by a car whatever the result of its attempts to park was, and by a car which finally acquires parking service. Let us denote

- $f_a(t)$ – probability density function of the time spent in the queue by a car that acquires parking service;
- $f_q(t)$ – probability density function of the time spent in the queue by a car that enters the system (this includes both cars that acquire parking service and those who renege).

Using the characteristic function method, one can find the first p.d.f:

$$f_a(t) = \frac{\lambda}{G - 1} e^{-(\mu+\alpha)} e^{\frac{\lambda}{\alpha}(1 - e^{-\alpha t})}.$$

Having the p.d.f. $f_a(t)$, we can calculate numerically the fraction of those who succeeded to park within $\tau$ minutes after arrival:

$$\psi_\tau = \int_0^\tau f_a(t) dt.$$

The probability distribution of the time spent in the queue regardless of the cruising outcome is related to the $f_a(t)$:

$$f_q(t) = P[A][f_a(t) - \alpha e^{-\alpha t} K(t)] + \alpha e^{-\alpha t},$$

$$K(t) = \frac{e^{\frac{\lambda}{\alpha}} \Gamma\left(\frac{\mu}{\alpha}\right)}{\left(\frac{\lambda}{\alpha}\right)^{\frac{\mu}{\alpha} - 1} (G - 1)} \left[I\left(\frac{\lambda}{\sqrt{\alpha\mu}}, \frac{\mu}{\alpha} - 1\right) - I\left(\frac{\lambda e^{-\alpha t}}{\sqrt{\alpha\mu}}, \frac{\mu}{\alpha} - 1\right)\right].$$

Having the p.d.f. $f_q(t)$, we can calculate the average waiting (cruising) time for those who found no free parking spot on arrival and were cruising for some positive amount of time:

$$W_\kappa = \int_0^\infty t f_q(t) dt = \frac{1 - P[A]}{\alpha}.$$

Finally, to get an unconditional waiting time, we simply multiply the above expression by the blocking probability:

$$w_\kappa = b \cdot W_\kappa.$$

**A2.4 Formulas related to the gamma function**

The following formulas for some finite sums and infinite series involving gamma function were used for the calculation of the probability states $p_k$ and average queue length $Q^*$:

$$\sum_{r=0}^{c} \frac{rx^r}{r!} = \frac{x[e^x \Gamma(c+1,x) - x^c]}{\Gamma(c+1)};$$

$$\sum_{r=0}^{\infty} \frac{1}{\Gamma(x+r)} = \frac{e(x-1)\gamma(x-1,1)}{\Gamma(x)};$$

$$\sum_{r=0}^{\infty} \frac{y^r}{\Gamma(x+r)} = \frac{e^y y^{1-x} \gamma(x-1,y)}{\Gamma(x-1)};$$

$$\sum_{r=0}^{\infty} \frac{ry^r}{\Gamma(x+r)} = \frac{xy^{1-x}[y^x + (y-x+1)e^y \gamma(x,y)]}{\Gamma(x+1)}.$$